\documentclass[epj,final]{svjour}
\usepackage{graphicx}
\usepackage{dcolumn}
\usepackage{bm}
\begin{document}
\title{Improved radiative corrections for $(e,e'p)$ experiments:
Beyond the peaking approximation and implications of the soft-photon approximation}
\author{Florian Weissbach, Kai Hencken, Daniela Rohe, Ingo Sick, 
Dirk Trautmann}
\institute{%
Departement f\"ur Physik und Astronomie,
Universit\"at Basel,
CH -- 4056 Basel, Switzerland}%
\date{\today}
\abstract{%
Analyzing $(e,e'p)$ experimental data involves corrections for
radiative effects which change the interaction kinematics and which have to
be carefully considered in order to obtain the desired accuracy.
Missing momentum and energy due to brems\-strah\-lung 
have so far often been incorporated into the simulations and
the experimental analyses using the peaking approximation.
It assumes that all brems\-strah\-lung
is emitted in the direction of the radiating particle.
In this article we introduce a full angular Monte Carlo simulation
method which overcomes this approximation.
As a test, the angular distribution of the brems\-strah\-lung photons
is reconstructed from H$(e,e'p)$ data. 
Its width is found to be underestimated by the peaking approximation
and described much better by the approach developed in this work.
The impact of the soft-photon approximation on the photon angular
distribution is found to be minor as compared to
the impact of the peaking approximation.
}
\authorrunning{F.~Weissbach {\em et al.}}
\titlerunning{Improved radiative corrections for $(e,e'p)$ experiments: 
Beyond the peaking approximation}
\maketitle
\section{Introduction}
\label{sec1}
Much of our knowledge about nuclear structure,
{\em e.g.}~the momentum distribution of nucleons in the nucleus,
is based on $(e,e'p)$ experiments.
Currently several such experiments are carried
out at the Thomas Jefferson National Accelerator Facility (TJNAF)
in Newport News and at the Mainz Microtron (MAMI) in Mainz,
taking data at high initial momenta and removal energies in particular.
These experiments are aiming at a deeper understanding {\em e.g.}~of
short-range correlations in nuclei and their results are used to check
important ingredients of modern many-body theories.\\

In $(e,e'p)$ experiments all particles involved
are subject to the emission of brems\-strah\-lung.
On the one hand, consideration of brems\-strah\-lung contributions
is necessary to renormalise the higher-order QED amplitudes.
In the second order,
divergences from the brems\-strah\-lung diagrams cancel with those 
resulting from vertex corrections
as has already been shown by Schwinger in 1949 \cite{schwinger}
for electrons scattering off an external potential
and for electron-proton scattering by Tsai in 1961 \cite{tsai},
including divergent contributions from the two-photon exchange
(TPE) diagrams.
On the other hand brems\-strah\-lung modifies the cross section
integrated over finite intervals of energy loss.
Brems\-strah\-lung photons can be so energetic that they influence 
the electron's and the proton's
three-momenta considerably; thereby also the momentum transfer between
the two particles is changed.
This phenomenon has been studied in a number of papers 
\cite{schiff,yennie,maximonisabelle0,maximonisabelle1,maximon,motsai,borie,borie2,friedrich,calan,aki1,florizone,maximontjon,afanasev1,afanasev2,makins,afanasev3}
both for inclusive and exclusive electron scattering 
experiments.\\

\begin{figure}[t]
\centering
\includegraphics[width=9cm]{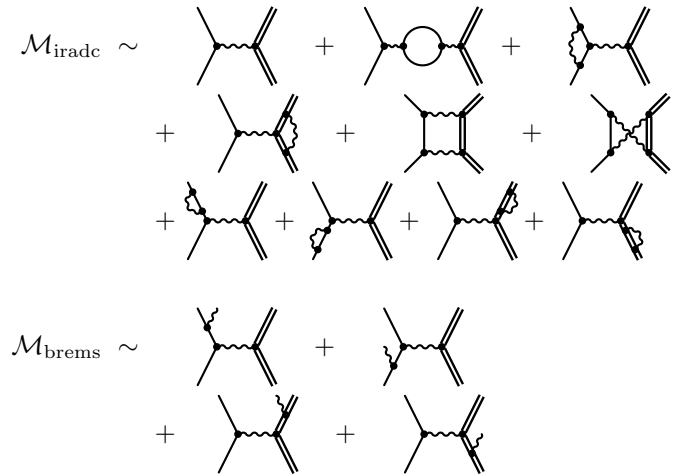}
\caption{\label{fig0} The Feynman diagrams contributing to the
internal radiative corrections together with the Born amplitude (top) 
and the diagrams contributing to the brems\-strah\-lung amplitude (below).
The label 'iradc' is short hand for {\em internal radiative corrections}
and 'brems' stands for {\em brems\-strah\-lung}.}
\end{figure}

Radiative corrections to $(e,e'p)$ scattering, including
brems\-strah\-lung, vertex corrections, and vacuum polarization
(see Feynman diagrams in fig.~\ref{fig0})
can in principle be calculated exactly in (pure) QED
and to a good accuracy also including hadronic loops.
But for practical purposes several approximations are usually employed
when correcting experimental data for radiative effects \cite{simc}.\\

One of them is the soft-photon approximation (SPA).
It makes use of the fact that in the limit where $\omega^0\rightarrow 0$
a brems\-strah\-lung photon with energy $\omega^0$ has neither a kinematic 
effect on the scattering process nor an effect on the QED propagators and 
amplitudes.
Then the SPA cross section factorizes into the elastic first-order Born
cross section times the probability for emitting a brems\-strah\-lung
photon with vanishing energy.
Analysis procedures for $(e,e'p)$ experiments make use of the SPA
\cite{simc} because it simplifies the calculation of multi-photon 
brems\-strah\-lung considerably \cite{makins}.\\

Multi-photon brems\-strah\-lung has to be included into electron
scattering data analysis \cite{makins,gupta} in order to 
both impose the physical asymptotic behaviour on the cross section 
and to achieve percent level accuracy.
While these higher-order brems\-strah\-lung contributions are, 
in principle, also computable exactly in QED, their evaluation
has to be truncated for practical purposes.
The SPA is convenient, because it allows for straight-forward inclusion
of multi-photon brems\-strah\-lung into data analysis
to {\em all} orders \cite{makins}, as we will see in sect.~\ref{sec2}.\\

The SPA is only valid in the limit of vanishing brems\-strah\-lung
photon energy; however in radiative correction procedures the SPA is applied to 
photons with energies of up to several hundred MeV.
The question arises up to which photon energies the SPA can be considered
a good approximation.
This paper will not answer that question albeit we will present
indications that one of the physical observables (the missing energy)
does exhibit sensitivity to shortcomings of the SPA.\\

$(e,e'p)$ data analyses do not employ the 'pure' SPA (called pSPA in
the remainder of this paper) described above.
Usually the SPA is modified such that it (i) takes into account 
kinematic effects due to emission of finite energy brems\-strah\-lung 
photons; and it (ii) evaluates the form factors at a modified
value of the momentum transfer of the virtual exchanged photon, $q$.
We will refer to this 'modified' SPA as the mSPA.
In mSPA each particle emitting brems\-strah\-lung is put onto
the mass shell.\\

The SPA neglects the proton structure at the brems\-strah\-lung vertex.
But, as has been shown in ref.~\cite{maximontjon}, for the 
kinematic settings considered in the present ma\-nu\-script, 
the influence of the proton structure at the brems\-strah\-lung vertex
is not important.\\

\begin{figure}
\centering
\includegraphics[height=7cm]{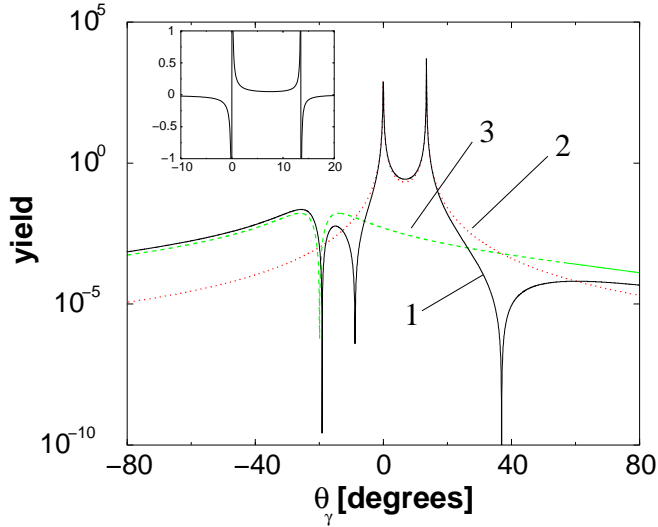}
\caption{\label{fig2} Single-photon angular distribution of 
brems\-strah\-lung using the SPA.
The full line (1) (black) shows the full angular distribution $A({\Omega_\gamma})$
from eq.~(\ref{a}), the dotted line (2) (red) considers pure electron contributions 
($ee$) only, the dashed line (3) (green) shows the pure proton contribution ($pp$).
The interference term is partially negative and is shown in
the inset graph. It is small compared to the other contributions.
The momentum transfer is $Q^2=15\,{\rm GeV}^2$, the entire kinematic
setting can be found in table~\ref{tab3}.
The dip for the proton is due to the fact that a massive
particle cannot radiate a photon in forward direction.
The same is true for electrons but the width of the respective
dip is extremely narrow \cite{makins,junior}.}
\end{figure}

\begin{figure}[t]
\centering
\includegraphics[height=8cm]{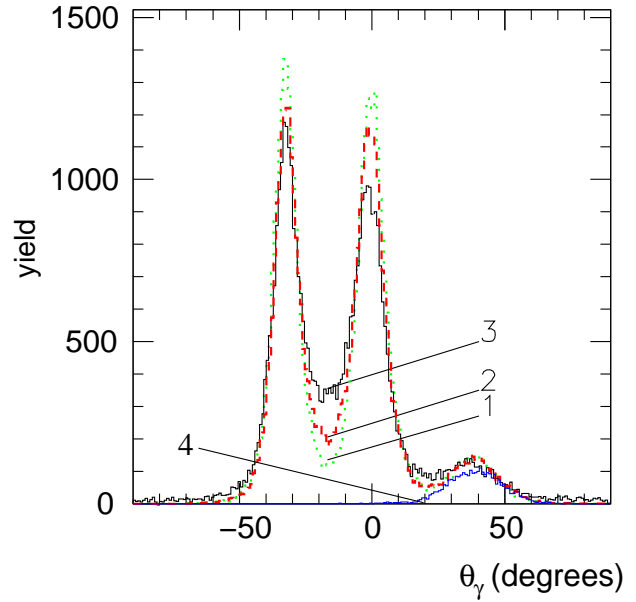}
\caption{\label{fig1} 
Brems\-strah\-lung angular distribution for the 
H$(e,e'p)$ reaction at the kinematics given in table~\ref{tab1}.
The solid curve (3) (black) shows the measured experimental
angular distribution of brems\-strah\-lung.
The experimental photon angle $\theta_\gamma$ is reconstructed from
the missing momentum according to eq.~(\ref{angles}).
The dotted line (1) (green) and the dashed line (2) (red) correspond to a Monte Carlo 
simulation based on the PA and take into account detector resolution.
The red line takes into account the emission of brems\-strah\-lung from
$e$, $e'$, and $p'$, whereas the green line only allows for
brems\-strah\-lung emitted solely from either $e$, $e'$, or $p'$.
The solid line (4) (blue) in the vicinity of the proton direction simulates a 
deficiency of the apparatus (proton punch-through effects).
The PA overestimates the two photon peaks in the two electron directions 
($e$ and $e'$) and it underestimates the brems\-strah\-lung between
these two peaks as well as between the peaks in
incident electron direction $e$ and in proton direction $p'$
and the radiation on the large-angle tail of the $e'$-peak.
Both data and simulations account for luminosities and detector efficiencies,
and no arbitrary normalization factor is needed.}
\end{figure}

The other approximation used in radiative correction procedures
is the peaking approximation (PA).
Most of the brems\-strah\-lung photons from the electron are emitted
either in the direction of the incoming ($e$) or outgoing electron ($e'$)
and one can observe two radiation peaks at the respective angles.
The proton ($p'$) brems\-strah\-lung is much less peaked.
At very high momentum transfers one can see a bump (rather than a peak) 
in its direction, too (see fig.~\ref{fig2}).
The PA, first proposed for $(e,e')$ experiments
by L.~I.~Schiff \cite{schiff} in 1952, makes use of this observation 
by assuming that {\it all} radiation goes either in the direction
of the incoming electron, or the scattered electron.
With the advent of coincidence experiments the PA
was extended to $(e,e'p)$ data \cite{makins}, assuming that the proton
brems\-strah\-lung was peaked, too.
The PA projects the non-peaked contributions to the brems\-strah\-lung 
photon angular distribution onto the three peaks.
Especially between the two radiation peaks due to electron brems\-strah\-lung
the discrepancy with data becomes large (see fig.~\ref{fig1}), limiting 
the accuracy of $(e,e'p)$ data analyses 
\cite{maximonisabelle0,maximonisabelle1}.\\

The purpose of this paper is to remove the PA from $(e,e'p)$ data 
analyses.
The need for the removal of the PA became evident when looking at the 
brems\-strah\-lung photon angular distribution in H$(e,e'p)$ experiments 
(see fig.~\ref{fig1}).
In this paper, we introduce a full angular Monte Carlo 
(FAMC) method which generates multi-photon brems\-strah\-lung events according 
to the mSPA photon angular distribution. 
A similar FAMC code for $(e,e'p)$ experiments has been described in 
ref.~\cite{afanasev3}.
But it has not been inserted into any data analysis codes nor
does it handle multi-photon brems\-strah\-lung.
In connection with virtual Compton scattering ref.~\cite{junior}
introduces a numerical calculation of radiative corrections beyond the PA,
but it considers single-photon emission only, whereas multi-photon
contributions are large.
To check our results against experimental data we use the
{\sc simc} analysis code \cite{simc} for Hall C at TJNAF
and E97-006 experimental data \cite{daniela}.\\

While we do not want to anticipate the results from sec.~\ref{sec7} at this 
stage we do state here on a preliminary basis that removing the PA can only 
be a first step on the way to an improved calculation not relying on 
the SPA.
For beam energies envisaged for the TJNAF upgrade, 
a calculation going beyond the SPA might become necessary, albeit
an exact multi-photon brems\-strah\-lung calculation is impracticable.\\

This paper is organized as follows: In sect.~\ref{sec2}
we introduce the brems\-strah\-lung cross section including multi-photon
brems\-strah\-lung, discussing the QED divergences.
Our calculation partially follows ref.~\cite{makins},
as the resulting equations form the basis for our FAMC calculation.
In sect.~\ref{sec4} we extend this approach to a FAMC
simulation allowing for any number of brems\-strah\-lung photons emitted
into the full solid angle according to the full angular distribution.
In sect.~\ref{sec5} we compare the results of the FAMC
\begin{table}
\begin{center}
\begin{tabular}{ll}\hline
$Q^2$ & $15\,{\rm GeV}^2$\\ \hline
$k^0$ & $21.00\,{\rm GeV}$\\ 
$|{\bf k'}|$ & $13.00\,{\rm GeV}$\\ 
$|{\bf p'}|$ & $8.882\,{\rm GeV}$\\ 
$\theta_{\rm e}$ & $13.5^o$\\ 
$\theta_{\rm p}$ & $-19.9^o$\\ \hline
\end{tabular}
\end{center}
\caption{\label{tab3} The kinematic setting used in fig.~\ref{fig2}.}
\end{table}
\begin{table}
\begin{center}
\begin{tabular}{ll}\hline
$Q^2$ & $2\,{\rm GeV}^2$\\ \hline
$k^0$ & $3.120\,{\rm GeV}$\\ 
$|{\bf k'}|$ & $2.050\,{\rm GeV}$\\ 
$|{\bf p'}|$ & $1.700\,{\rm GeV}$\\ 
$\theta_{\rm e}$ & $32.5^o$\\ 
$\theta_{\rm p}$ & $-38.5^o$\\ \hline
\end{tabular}
\end{center}
\caption{\label{tab1} The kinematic setting at which the standard
radiative corrections and the FAMC simulation are compared to data.}
\end{table}
simulation to the PA using the {\sc simc} code, and in sect.~\ref{sec7} 
we discuss scope and validity of the SPA, comparing it to the exact
QED calculation for single-photon brems\-strah\-lung from the electron
(which will be called '$1\gamma$ calculation').\\

The speed of light has been set to $c=1$ throughout the paper.
\section{Brems\-strah\-lung cross section}
\label{sec2}
\setlength{\unitlength}{1mm}
In order to obtain the electron-proton cross section to order $\alpha^2$ 
including brems\-strah\-lung with energy less than $\omega_{\rm max}$,
\begin{eqnarray}
\label{cross1}
\frac{d\sigma}{d\Omega_{\rm e}}(\omega^0<\omega_{\rm max}) \, ,
\end{eqnarray}
where $\omega^0$ is the brems\-strah\-lung photon energy,
the amplitudes depicted in fig.~\ref{fig0} are considered.
The four brems\-strah\-lung diagrams contributing to ${\cal M}_{\rm brems}$ 
are divergent in the limit of vanishing brems\-strah\-lung photon energy
$\omega^0$.
These divergences cancel the ones both from the TPE 
diagrams\footnote{The divergences from the TPE diagrams cancel with the one
from the electron-proton brems\-strah\-lung interference term which
appears after squaring the full scattering amplitude.}
and the vertex corrections \cite{schwinger}.
The TPE diagrams are special cases.
While consideration of their divergent pieces is necessary
in order to remove all divergences from the scattering amplitudes,
their finite contributions are known to be negligible in electron
scattering experiments \cite{lewis,drellruderman,drellfubini,campbell}
unless a very small $L(T)$-contribution is determined via an $LT$-separation.
Mo and Tsai calculated the TPE diagrams approximately using only the nucleon
intermediate state in the limit where one of the two
exchanged photons has zero momentum. 
They applied this approximation both in the
numerator and in the denominator of the fermion propagator 
\cite{motsai}.
Maximon and Tjon improved this calculation by removing this 
approximation from the denominator of the fermion propagator
\cite{calan,maximontjon}.
And Blunden, Melnitchouk, and Tjon did the calculation using the
full propagator \cite{blunden03,guichon03}.
According to ref.~\cite{blunden03} a model-dependent calculation
of the influence of the TPE yields effects of the order
of 1-2\% for the kinematic settings considered in the present
paper.
Most $(e,e'p)$ analysis codes follow the calculation by Mo and Tsai
\cite{makins,simc}.\\

The SPA allows us to approximate the four brems\-strah\-lung diagrams 
by a product of the Born amplitude times a correction factor.
In SPA, {\em e.g.}~the amplitude for incident electron brems\-strah\-lung can 
be approximated as
\begin{eqnarray}
\label{spa}
{\cal M}_{\rm ei} = e {\cal M}_{\rm ep}^{(1)}
\left(-\frac{\varepsilon\cdot k}{\omega\cdot k}\right) 
\hspace{4mm}(\omega^0\rightarrow 0) \, .
\end{eqnarray}
This amplitude corresponds to the second Feynman dia\-gram of 
${\cal M}_{\rm brems}$ in fig.~\ref{fig0}.
${\cal M}_{\rm ep}^{(1)}$ is the first-order Born
amplitude, $\omega=\omega^0(1,1,\Omega_\gamma)$ the photon four-momentum,
$\varepsilon$ is the brems\-strah\-lung photon helicity vector,
and $k=(k^0,{\bf k})$ the incident electron's four-momentum.
The four-momentum of the scattered electron will be denoted as
$k'=(k'^0,{\bf k'})$, and for the proton we will use $p=(p^0,{\bf p})$ 
(incoming) and $p'=(p'^0,{\bf p}')$ (outgoing).\\

Evaluating the Feynman diagrams in the SPA
one can show \cite{makins} that the cross section
for single-photon brems\-strah\-lung is
\begin{eqnarray}
\label{cross2}
\frac{d\sigma}{d\Omega_{\rm e} d\Omega_\gamma d\omega^0}
=\frac{d\sigma^{(1)}}{d\Omega_{\rm e}}\frac{A({\Omega_\gamma})}{\omega^0} \, ,
\end{eqnarray}
where ${\Omega_\gamma}$ denotes the brems\-strah\-lung photon angles
and $d\sigma^{(1)}/d\Omega_{\rm e}$ is the Born cross section.
In the cross section (\ref{cross2}) the dependences on photon energy and 
photon angle factorize and
\begin{eqnarray}
\label{a}
A({\Omega_\gamma})\equiv\frac{\alpha\omega^{0^2}}{4\pi^2}
\left(
-\frac{k'}{\omega\cdot k'} 
+\frac{p'}{\omega\cdot p'}
+\frac{k} {\omega\cdot k}
-\frac{p} {\omega\cdot p}
\right)^2 
\end{eqnarray}
does not depend on the photon energy.\\

Integrating over photon angles and energies the total cross section 
for emitting a photon with energy smaller than $\omega_{\rm max}$ can
be written as
\begin{eqnarray}
\label{cross3}
\frac{d\sigma}{d\Omega_{\rm e}}(\omega^0<\omega_{\rm max})
&=&\frac{d\sigma^{(1)}}{d\Omega_{\rm e}}
[1-\delta_{\rm brems}(\omega_{\rm max})-\delta_{\rm iradc}]\, .
\nonumber \\
&&
\end{eqnarray}
The necessary integration techniques can be found in \cite{thooft},
the remaining calculations are explicitly carried out in ref.~\cite{makins}.
The contributions from vertex correction and vacuum polarization 
(the internal radiative corrections) are included in
\begin{eqnarray}
\label{deltahard}
\delta_{\rm iradc}\equiv 2\alpha
\left[-\frac{3}{4\pi}\log\left(\frac{Q^2}{m^2}\right)+\frac{1}{\pi} 
-\sum_i \delta_i^{\rm vp}(Q^2)\right] \, ,
\end{eqnarray}
where the vacuum polarization contribution is
\begin{eqnarray}
\delta_i^{\rm vp}(Q^2)\equiv
\frac{1}{3\pi}
\left[
-\frac{5}{3}+\log\left(\frac{Q^2}{m_i^2}\right)
\right]
\end{eqnarray}
in the ultra-relativistic (UR) limit. 
This expression does not only contain electron-positron
loops but also heavier lepton and light quark-anti-quark loops,
$m_i$ denoting their respective mas\-ses.
The brems\-strah\-lung is contained in
\begin{eqnarray}
\label{deltasoft}
\delta_{\rm brems}(\omega_{\rm max})
&\equiv&
\frac{\alpha}{\pi}
\left\{
\log\left(\frac{|{\bf k}||{\bf k}'|}{\omega_{\rm max}^2}\right)
\left[\log\left(\frac{Q^2}{m^2}\right)-1
\right]
\right.\nonumber\\&&
+\log\left(\frac{p^0 p'^0}{\omega_{\rm max}^2}\right)
\left[
\log\left(\frac{Q^2}{M^2}\right)-1
\right]
\nonumber\\
&&
+\frac{1}{2}\log^2\left(\frac{p'^0}{M}\right)
+\,\log\left(\frac{p^0 p'^0}{\omega_{\rm max}^2}\right)
\nonumber\\
&&
\times
\log\left(\frac{|{\bf k}|}{|{\bf k}'|}\right)
+\log\left(\frac{|{\bf k}||{\bf k}'|}{\omega_{\rm max}^2}\right)
\log\left(\frac{|{\bf k}|}{|{\bf k}'|}\right)
\nonumber\\
&&\left.
+\frac{1}{2}\log\left(\frac{|{\bf k}||{\bf k}'|}{M^2}\right)
\log\left(\frac{|{\bf k}|}{|{\bf k}'|}\right)
\right\} \, ,
\end{eqnarray}
also given in the UR limit.
The single-photon cross section (\ref{cross3}) is still divergent
in the limit of vanishing $\omega_{\rm max}$.
By taking into account higher-order brems\-strah-lung (multi-photon
brems\-strah\-lung) this divergency is rendered finite 
\cite{yennie,gupta} and, at the same time, experimental accuracy is
enhanced.
It was first shown in ref.~\cite{yennie} that in fact {\it all}
orders of brems\-strah\-lung contributions can be considered by just
exponentiating the brems\-strah\-lung term in the cross section 
(\ref{cross3}), yielding
\begin{eqnarray}
\label{cross4}
\frac{d\sigma}{d\Omega_{\rm e}}(\omega_i^0<\omega_{\rm max})
&=&\frac{d\sigma^{(1)}}{d\Omega_{\rm e}}
\exp[-\delta_{\rm brems}(\omega_{\rm max})][1-\delta_{\rm iradc}]\, .
\nonumber
\\&&
\end{eqnarray}
The index $i$ indicates that an infinite number of photons, 
each with an energy less than $\omega_{\rm max}$, is emitted.
Exponentiating $\delta_{\rm brems}$ leads to the correct asymptotic behaviour 
of the cross section (\ref{cross3}) as $\omega_{\rm max}\rightarrow 0$.\\


We now consider the cross section for emitting $n$ photons with
an energy larger than an artifically introduced energy cut-off
parametet $E_{\rm min}$ together with multi-photon emission of photons
with individual energies less than the energy cut-off $E_{\rm min}$
\cite{makins},
\begin{eqnarray}
\label{cross5}
\frac{d\sigma(n,E_{\rm min})}{d\Omega_{\rm e} d\omega_1^0 
d\Omega_1 ... d\omega_n^0 d\Omega_n}
&=&\frac{d\sigma^{(1)}}{d\Omega_{\rm e}}\exp[-\delta_{\rm
  brems}(E_{\rm min})]
\nonumber\\
&&
\times
(1-\delta_{\rm iradc})
\nonumber\\
&&
\times
\frac{1}{n!}
\frac{A({\Omega_1})}{\omega_1^0} ...
\frac{A({\Omega_n})}{\omega_n^0}
\nonumber\\
&&
\times
\theta(\omega_1^0-E_{\rm min})...\theta(\omega_n^0-E_{\rm min}) \, .
\nonumber\\
&&
\label{eq:differential}
\end{eqnarray}
In order to integrate this cross section, let us introduce an ``acceptance function''
$\chi_A^n(\Omega_e,\omega_1^0,\Omega_1,\ldots,\omega_n^0,\Omega_n)$, 
depending on the kinematical variables of the scattered electron and
the $n$ photons with energies larger than $E_{\rm min}$.
Multiplying eq.~({\ref{eq:differential}) with $\chi_A^n$, integrating over all photon
energies up to an upper boundary $E_{\rm max}>E_{\rm min}$, chosen
large enough to include all photons, where $\chi_A^n$
is non-zero, we obtain the cross section
\begin{eqnarray}
\label{cross6}
\frac{d\sigma}{d\Omega_{\rm e}}\left[\chi_A\right]
&=&\frac{d\sigma^{(1)}}{d\Omega_{\rm e}}\exp[-\delta_{\rm
  brems}(E_{\rm min})]
(1-\delta_{\rm iradc})
\nonumber \\
&&\times\sum_{n=0}^\infty\frac{1}{n!}
\prod_{i=1}^n 
\int_{E_{\rm min}}^{E_{\rm
    max}}\frac{d\omega_i^0}{\omega_i^0}
\int d\Omega_i A({\Omega_i}) \chi_A^n
\, .
\nonumber\\
&&
\end{eqnarray}
This way, cross sections with very general restrictions on, {\em e.g.},~the 
energy loss by the photons can be calculated. 
The acceptance function $\chi_A\le 1$ 
is the probability for the event to be counted in 
$\frac{d\sigma}{d\Omega_{\rm e}}\left[\chi_A\right]$.\\

Proceeding towards a FAMC calculation we evaluate the cross section
(\ref{cross6}) using Monte Carlo integration techniques. 
In order to rewrite the integral into the standard Monte Carlo sum over randomly
selected values, we need to rewrite the cross section in terms of probability 
density functions (PDFs) for the photon variables. 
The PDFs are then used to generate the random values.\\

As we want to keep the shape of $\chi_A^n$ general, it is kept
in the expression for the cross section and is implemented in the Monte Carlo 
generator by a standard rejection algorithm. 
The integral is easily converted into the shape demanded by the Monte Carlo
generation by multiplying and dividing it with its value for $\chi_A^n=1$.\\

The value of the angular integral (without any $\chi_A$)
\begin{eqnarray}
\label{lambda}
\lambda\equiv\int d\Omega_\gamma A({\Omega_\gamma})\, 
\end{eqnarray}
is independent of the photon energy and $A({\Omega_\gamma})$ is the 
angular distribution from eq.~(\ref{a}) which is plotted in fig.~\ref{fig2}
(for $\phi=0$) for a sample kinematic configuration (see table \ref{tab3}).
The integrals over the $\omega_i^0$ can be trivially solved
and the product over $i$ just yields a power of $n$.\\

Re-writing the cross section in terms of the  PDFs for the brems\-strah\-lung 
photon energies $\omega_i^0$, the angular distribution and their
multiplicity $n$, leads to
\begin{eqnarray}
\label{cross10}
\frac{d\sigma}{d\Omega_{\rm e}}\left[\chi_A \right]
&=&\frac{d\sigma^{(1)}}{d\Omega_{\rm e}}
\exp[-\delta_{\rm brems}(E_{\rm min})](1-\delta_{\rm iradc}) 
\nonumber \\
&&
\times
e^{\lambda\log\left(\frac{E_{\rm max}}{E_{\rm min}}\right)}
\sum_{n=0}^\infty 
{\rm PDF}(n)
\nonumber\\
&&
\times\prod_{i=1}^n \int d\omega_i^0 {\rm PDF}(\omega_i^0)
\nonumber\\
&&
\times
\int d\Omega_i {\rm PDF}(\Omega_i) \chi_A^n
\, ,
\end{eqnarray}
where
\begin{eqnarray}
\label{pdf1}
{\rm PDF}(\omega_i^0)\equiv 
\frac{1}{\omega_i^0 \log\left(\frac{E_{\rm max}}{E_{\rm min}}\right)} \, 
\end{eqnarray}
is the PDF for the photon energies $\omega_i^0$,
\begin{equation}
\label{pdf3}
{\rm PDF}(\Omega_i)\equiv\frac{A(\Omega_i)}{\lambda}
\end{equation}
is the one for the angular distribution; and
\begin{eqnarray}
\label{pdf2}
{\rm PDF}(n)\equiv\frac{1}{n!}
\left[\lambda \log \left(\frac{E_{\rm max}}{E_{\rm min}}\right)\right]^n
e^{-{\lambda}\log\left(\frac{E_{\rm max}}{E_{\rm min}}\right)} \, ,
\end{eqnarray}
is the PDF for the photon multiplicity, which is just a Poisson distribution.
The total cross section (\ref{cross10}) does not depend on $E_{\rm min}$ because 
$\delta_{\rm brems}(E_{\rm min})$ and $\lambda$ are such that $E_{\rm min}$ 
cancels, as has been shown by R.~Ent {\em et al.}~in ref.~\cite{makins}.\\

In the same reference \cite{makins} it is shown that the cross section for 
emitting several photons each with energy less than a cut-off $E_{\rm min}$,
\begin{eqnarray}
\label{ente}
\frac{d\sigma}{d\Omega_{\rm e}}(\omega_i^0<E_{\rm min}) \, ,
\end{eqnarray}
is, within a correction of order $\alpha^2$, the same
for the case where instead of the individual photon energies
the {\em sum} of the energies of all brems\-strah\-lung
photons is smaller than the cut-off,
\begin{eqnarray}
\label{cross4b}
\frac{d\sigma}{d\Omega_{\rm e}}\left(\sum_i\omega_i^0<E_{\rm min}\right)
&=&
\frac{d\sigma}{d\Omega_{\rm e}}(\omega_i^0<E_{\rm min})
[1+{\cal O}(\alpha^2)] \, .
\nonumber\\
&&
\end{eqnarray}
Therefore the cross section (\ref{cross10}), within an order $\alpha^2$ 
correction, can be regarded as the cross section for multi-photon emission below a 
small cut-off $E_{\rm min}$, the {\em sum} of these soft photons being
$E_{\rm min}$, along 
with the emission of $n$ hard photons with energies above $E_{\rm min}$
and below $E_{\rm max}$. The dependence on $E_{\min}$ cancels \cite{makins}.
For practical purposes, in $(e,e'p)$ data analyses, $E_{\min}$ is
often set to a value below the
detector resolution, and the brems\-strah\-lung photons below
that small cut-off are not considered in the analyses, as even their
sum will not affect the measured result:
Eq.~(\ref{cross4b}) ensures that neglect of these photons only amounts to
missing energies below the detector resolution within an order $\alpha^2$
correction. For the value of $E_{\rm max}$ we can always use the total
energy of the incoming electron, as a reasonable
acceptance function disallows events, where the photons have
together more energy than the total energy available.\\

Starting with the cross section in (\ref{cross5}) we can also obtain
differential cross sections. 
For example, the cross section differential in the total energy of all emitted 
brems\-strah\-lung photons $E_{\rm tot}$,
%
$\frac{d\sigma}{d\Omega_{\rm e} dE_{\rm tot}} \, ,$
%
is calculated by choosing $\chi_A^n=\delta(\sum_{i=1}^n \omega_i^0 -
E_{\rm tot}^0)$, yielding
\begin{eqnarray}
\label{cross11}
\frac{d\sigma}{d\Omega_{\rm e} dE_{\rm tot}}
&=&\frac{d\sigma^{(1)}}{d\Omega_{\rm e}}
\exp[-\delta_{\rm brems}(E_{\rm min})](1-\delta_{\rm iradc}) 
\nonumber \\
&&
\times
e^{\lambda\log\left(\frac{E_{\rm max}}{E_{\rm min}}\right)}
\sum_{n=0}^\infty 
{\rm PDF}(n)
\nonumber \\
&&
\times
\prod_{i=1}^n \int d\omega_i^0 {\rm PDF}(\omega_i^0)
\nonumber \\
&&
\times
\delta\left(\sum_{i=1}^n\omega_i^0-E_{\rm tot}\right)
\, ,
\end{eqnarray}
where we have made use of the fact that in this case the integration over the
angular variables can be done trivially.\\

In the Monte Carlo simulation events are generated according to
the PDFs (\ref{pdf1}), (\ref{pdf3}), and (\ref{pdf2}) and the results
are binned in the vicinity of
\begin{eqnarray}
\label{binning}
\sum_i \omega_i^0 \approx E_{\rm tot} \, .
\end{eqnarray}

As discussed above, $n$, $\Omega_i$, and $\omega_i^0$ have to be generated according 
to the PDFs (\ref{pdf1}) to (\ref{pdf2}).
While this approach has been used in $(e,e'p)$ data analysis codes
(in connection with the PA) and is well-established \cite{simc}, an 
$(e,e'p)$ data analysis code using a multi-photon FAMC simulation is novel.
In the next section we will describe how the photon angular distribution
(\ref{a}) is generated.
\section{Full angular Monte Carlo simulation}
\label{sec4}
For H$(e,e'p)$ data
the PA exhibits its limitations especially in the middle between
the two radiation peaks in $e$ and $e'$ directions\footnote{Only for 
H$(e,e'p)$ the missing-momentum vector and thus its {\em direction}
is solely due to emission of brems\-strah\-lung photons below pion threshold.}
where it underestimates the strength of the brems\-strah\-lung.
The same is true for the region between the peaks in $e$ and $p'$ directions.
In this section we introduce a FAMC simulation for multi-photon 
brems\-strah\-lung following the SPA distribution in eq.~(\ref{a}) in 
order to see whether this cures the problem.
In order to generate Monte Carlo events according to the angular
distribution (\ref{a}) we need a set of invertible envelope curves
${\hat A}_i(\Omega_\gamma)$ which limit $A(\Omega_\gamma)$ from above,
\begin{eqnarray}
\label{envelope}
\sum_i{\hat A}_i(\Omega_\gamma)\ge A(\Omega_\gamma)
\end{eqnarray}
for all photon angles $\Omega_\gamma$.
In order to obtain the exact distribution (\ref{a}) from the envelope
curves we then employ a standard rejection algorithm.\\

The envelope curve chosen by us consists of four terms. 
In order to be able to apply the mSPA we need to assign each
brems\-strah\-lung photon to one of the particles.
Three of the four contributions to the envelope curve can unambiguously be 
assigned to brems\-strah\-lung from the incoming and outgoing electron and 
the outgoing proton.
The fourth envelope term takes up the remaining part.
\begin{figure}
\centering
\includegraphics[height=8cm]{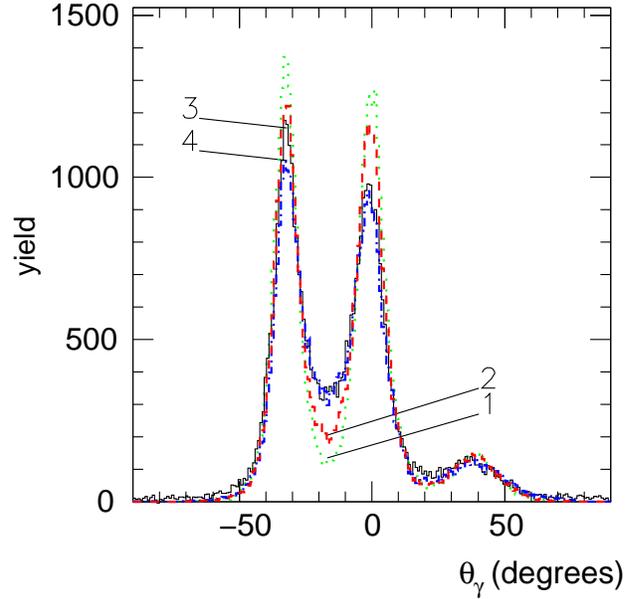}
\caption{\label{fig6} Angular distribution of brems\-strah\-lung photons.
The dotted line (1) (green) and the dashed line (2) (red) represent the simulations
using the PA (the difference between the two curves is explained in the 
caption to fig.~\ref{fig1}), and the solid line (3) (black) shows the measured 
(reconstructed) experimental angular distribution, as in fig.~\ref{fig1}.
The dot-dashed curve (4) (blue) represents our FAMC simulation (not using the PA; employing
method 3 described in the text; see fig.~\ref{fig11}).
The peaks in $e$ and in $e'$ direction generated by the FAMC
simulation (4) are broader than the ones from the PA, (1) and (2), 
and reproduce the data more accurately, especially between the peaks
in $e$ and $e'$ direction.
The height of the $e'$ peak is slightly underestimated by the
FAMC simulation (4), as well as the large-angle tail
beyond the $e'$-peak.
For details concerning the proton direction see fig.~\ref{fig6detail}.
As in fig.~\ref{fig1} both data and simlations are corrected for luminosities 
and detector efficiencies, and no arbitrary normalization was introduced.}
\end{figure}
\begin{figure}
\centering
\includegraphics[height=8cm]{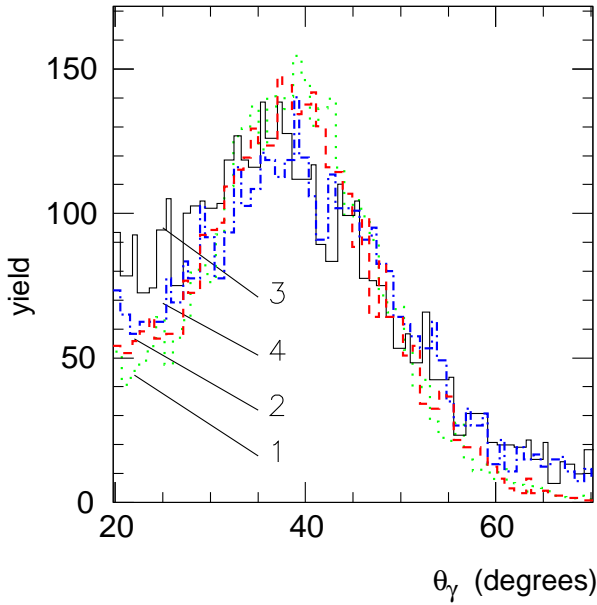}
\caption{\label{fig6detail} 
Angular distributions of brems\-strah\-lung photons.
This figure is a cutout of fig.~\ref{fig6}, focussing on the region around 
the proton direction.
On the small-angle tail of the proton peak the FAMC simulation (4) does 
not entirely overcome the gap between simulation and data.
However, the region around the proton direction is obscured by 
punch-throughs.}
\end{figure}
It is an angle-independent distribution at first but shaped by the rejection
algorithm into a contribution which is given by $ee$ interference.
There are several 'coin toss' methods to choose whether
an event created from the interference term is assigned to the incoming or the
outgoing electron or to both.
We employed three different ways of dealing with the interference term,
leading to slightly different results.
Together with the Monte Carlo photon energy generation (\ref{pdf1})
and with the photon multiplicity generation (\ref{pdf2}) each of these
three ways of dealing with the interference term constitutes a Monte
Carlo event generation method for the interference term.
\begin{enumerate}
\item The interference term (being essentially a function of the photon
angle $\theta_\gamma$)
is split into two parts, the 'left part' consisting of events
with angles closer to $\theta_{\rm e}$ and the 'right' part with
angles closer to zero.
Events closer to the incoming electron direction ('right') 
were counted for
the incoming electron whereas events closer to the outgoing electron
('left') direction were counted for the latter one.
\item In addition to method (1) the energy loss generated using (\ref{pdf1})
is randomly split between incident and scattered electron.
\item The emitted photon is randomly assigned to either
the incoming or the outgoing electron.
\end{enumerate}
For the final comparison between the standard brems\-strah\-lung treatment 
(using PA) and our FAMC simulation we used the third method
as it fitted the reconstructed photon distribution most accurately, as we 
will see in the next section.\\

Once a brems\-strah\-lung event has been assigned either to
the incident electron, the scattered electron or to the struck
proton, in mSPA the four-momenta, $k$, $k'$, and $p'$ are replaced by 
$k\rightarrow k-\omega$,
$k'\rightarrow k'-\omega$, or by
$p'\rightarrow p'-\omega$,
respectively, and the momentum transfer $q^2$ is adjusted and inserted into the form factors.
$\omega$ is the four-vector of the brems\-strah\-lung photon.\\

To check the results produced with our Monte Carlo routine against
experimental data we implemented it into the {\sc simc} code
\cite{simc} developed for Hall C at TJNAF.
We used a modified version which was used for the 
E97-006 $(e,e'p)$ experiment \cite{daniela}.
Computation times with and without the new FAMC simulation were similar.
\begin{figure}
\centering
\includegraphics[height=8cm]{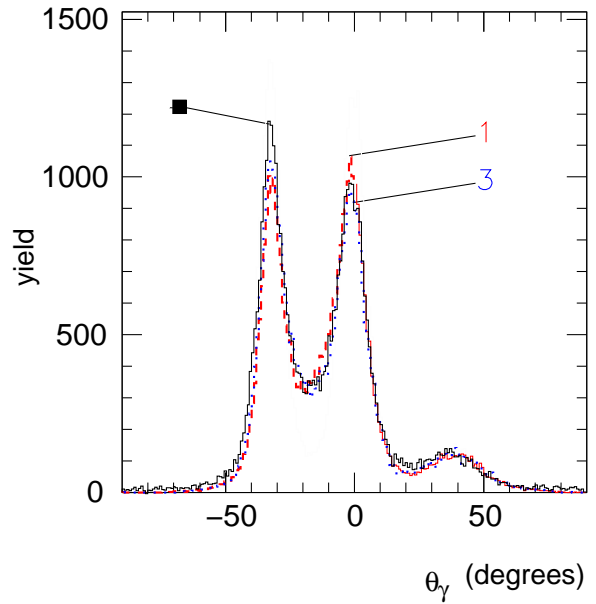}
\caption{\label{fig11} Photon angular distribution for the three different 
treatments of the interference term (see sec.~\ref{sec4}).
Method (1) is represented by the dashed line line (red)
and it coincides with method (2) (not shown).
The dotted line (blue) represents method (3), the solid line (black, marked with
the black square) shows the data.
Method (3) is found to reproduce the data most
accurately and is hence used for the FAMC simulation (4) in fig.~\ref{fig6}.}
\end{figure}
\begin{figure}
\centering
\includegraphics[height=8cm]{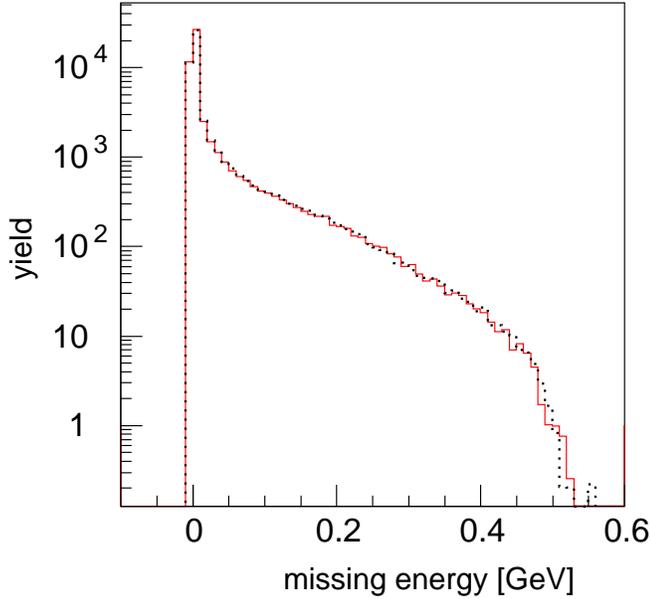}
\caption{\label{fig7} Logarithmic plot of missing-energy distribution.
The dotted line (black) was obtained using the PA,
the solid line (red) shows the results obtained with the FAMC simulation. 
The total radiated energy simulated for the latter
case is only about 0.3\% smaller than the one from the PA.
Reasons for this deviation are given at the end of sec.~\ref{sec5}.}
\end{figure}
\section{Results}
\label{sec5}
To test our approach we chose H$(e,e'p)$ kinematics with a beam energy
of $3.12\, {\rm GeV}$. 
We usually generated 600,000 successful Monte Carlo events per run to compare
PA and FAMC simulation.
Figs.~\ref{fig6} -- \ref{fig10} show results for the kinematic setting 
given in table {\ref{tab1}}.\\

The photon angles shown in figs.~\ref{fig1} and \ref{fig6} are obtained 
according to the prescription
\begin{eqnarray}
\label{angles}
\theta_\gamma
= \arctan\left(\frac{p_{{\rm m}_y}}{p_{{\rm m}_z}}\right) \, ,
\end{eqnarray}
where $p_{{\rm m}_y}$ and $p_{{\rm m}_z}$ are the missing momenta in $y$ and 
in $z$ direction, respectively.
Our co-ordinate system is the one used by SIMC, described in 
\cite{manual}.\\

As pointed out in the introduction, the PA underestimates non-peaked 
radiation especially between the radiation peaks in the directions of 
the incident and the scattered electron 
\cite{maximonisabelle0,maximonisabelle1}.
One can see in fig.~\ref{fig6} that the photon angular distribution
broadens when employing the FAMC simulation.
The gap between the experimentally determined brems\-strah\-lung distribution 
and PA (see fig.~\ref{fig1}) between the two radiation peaks in $e$ and $e'$
direction is filled.\\

When calculated with our FAMC method, also the peak in the proton direction 
fits the reconstructed brems\-strah\-lung data (see fig.~\ref{fig6detail}).
However, this has to be put into perspective as the proton brems\-strah\-lung 
is obscured by a detector related artefact (punch-through effects)
such that one cannot make a clear statement on the accuracy here.\\

For the kinematic setting shown in table \ref{tab1} the $ee$ interference
term discussed in the previous section was treated with method (3). 
This led to the best agreement with data as can be seen in fig.~\ref{fig11}.
The other two methods also improved the angular distribution 
of the brems\-strah\-lung but exhibited a slightly larger deviation
from the data concerning the amplitudes of the $e$ and $e'$ peaks.
At the kinematic setting shown in table \ref{tab1} 
the $ee$ interference term accounted for roughly 20\% of all events.\\

\begin{figure}
\centering
\includegraphics[height=8cm]{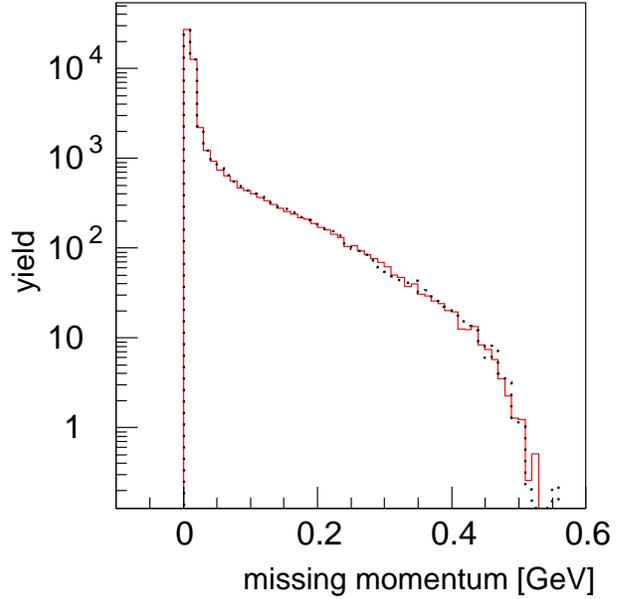}
\caption{\label{fig8} Logarithmic plot of the reconstructed missing-momentum
distribution.
The missing momentum is almost unaltered.
The FAMC simulation is represented by the solid line (red), the PA by the dotted 
line (black).}
\end{figure}

Looking at the missing-energy distribution in fig.~\ref{fig7} which
includes detector resolution and acceptances, we see that the total FAMC
yield is 0.3\% smaller than predicted by the PA, while a calculation
not taking into account detector resolution and acceptances would yield
identical results for FAMC and PA.
The 0.3\% difference is well within the systematic uncertainty usually 
attributed to the radiation correction.
The missing momenta (see fig.~\ref{fig8}) generated by the PA and by our 
FAMC code do hardly differ either, as the missing energy distribution.
The momentum distributions of electron and proton for the kinematics 
shown in table \ref{tab1} are also not changed significantly
by the FAMC calculations, as can be seen in figs.~\ref{fig9} and 
\ref{fig10}.\\

As a further check we also looked at kinematic settings with both
larger and smaller values of $Q^2$, while we let the beam energy
unaltered.
We compared again the FAMC simulation with the standard radiation code.
\begin{figure}
\centering
\includegraphics[height=8cm]{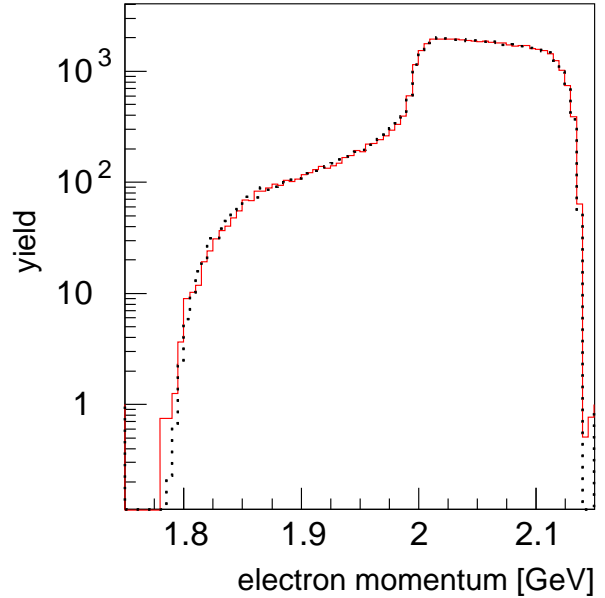}
\caption{\label{fig9} Electron momentum distribution.
The dotted line (black) was obtained using the PA,
the solid line (red) shows the results using the FAMC simulation.}
\end{figure}
Looking at the total yield in the acceptance we found differences of up 
to 3.0\%, the yield of the FAMC simulation usually being smaller than the 
standard analysis yield when going to higher momentum transfers and larger 
for small values of $Q^2$, as can be seen in table \ref{tab2} and in 
fig.~\ref{newfig}.\\

The differences in the total yield shown in table~\ref{tab2}, in 
figs.~\ref{fig7} and \ref{newfig} are related to the inappropriate 
application of the SPA.
It only shows up when including detector simulations into the data analysis.
\begin{figure}
\centering
\includegraphics[height=8cm]{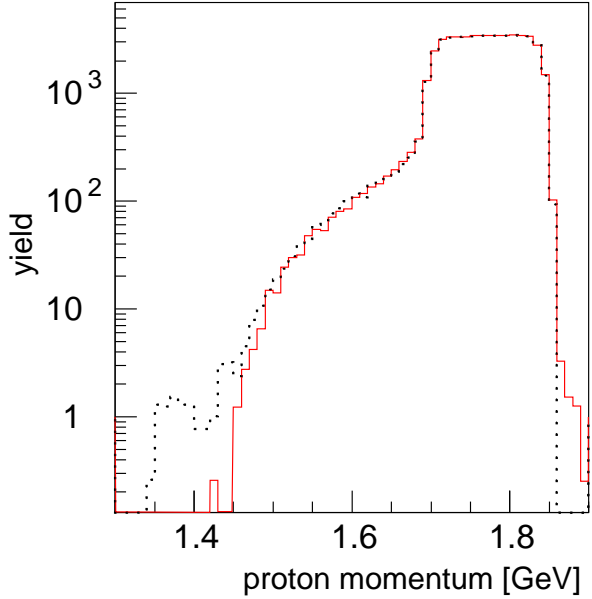}
\caption{\label{fig10} Proton momentum distribution.
FAMC simulation (solid line) (red) and peaking calculation (dotted line) (black) almost 
coincide.}
\end{figure}
\begin{table}
\begin{center}
\begin{tabular}{llllll}\hline
$Q^2/$GeV$^2$     & 0.61$^\star$ & 1.00$^\star$ & 2.00    & 3.00$^\star$ & 4.00$^\star$
\\ \hline
$|{\bf p'}|/$GeV  & 0.852        & 1.13    & 1.70        & 2.36    & 2.92\\
$|{\bf k'}|/$GeV  & 2.74         & 2.59    & 2.05        & 1.52    & 0.99\\ 
$\Delta$ yield    & +2.5\%       & +0.4\%  & --0.3\%     & --1.5\% & --3.0\% \\ \hline
\end{tabular}
\end{center}
\caption{\label{tab2} The differences in the yield between the standard 
analysis code and the FAMC simulation, integrated up to $0.7\, {\rm GeV}$.
The incident electron's energy is $k^0=3.120\,{\rm GeV}$ for all kinematic 
settings.
Reasons for the differences are given at the end of sec.~\ref{sec5}.
Missing energy plots for the kinematics settings marked with a star are
shown in fig.~\ref{newfig}.}
\end{table}
\begin{figure}
\centering
\includegraphics[height=14cm]{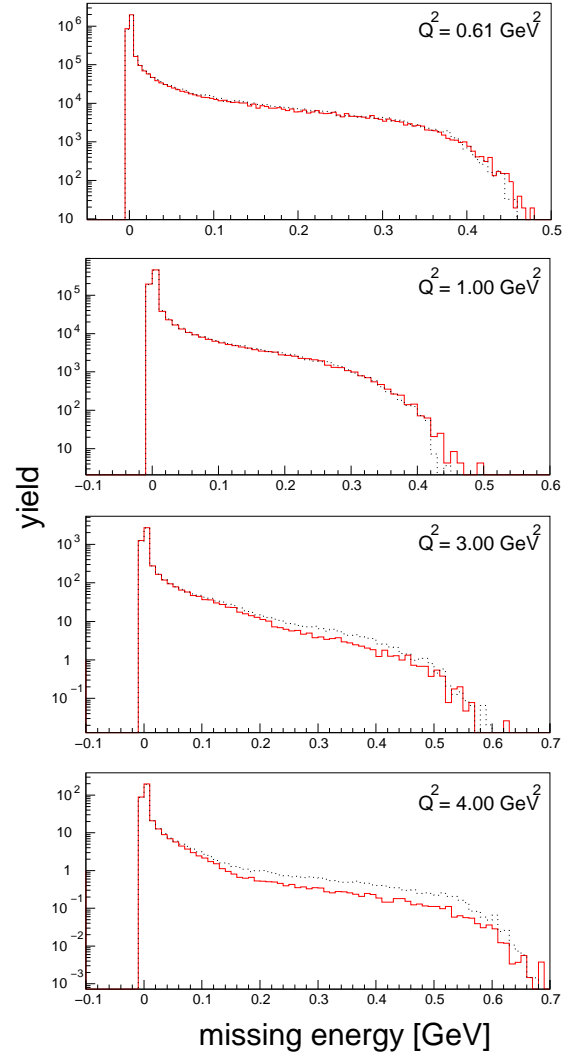}
\caption{\label{newfig} 
Logarithmic plots of missing-energy distributions for the remaining kinematic 
settings shown in table \ref{tab2}. 
The distribution for $Q^2=2.00\,{\rm GeV}^2$ has already been shown in 
fig.~\ref{fig7}.
The solid curves (red) represent the FAMC calculation, the dotted curves (black) show the 
PA calculation.}
\end{figure}
Our FAMC approach is more sensitive to problems caused by the SPA 
than the PA at certain kinematic settings.
It uncovers a problem of the SPA which is suppressed by the PA.
Including the full angular dependence of brems\-strah\-lung photons
(other than the trivial angular dependence of the PA)
can sometimes lead to energy gains for both electron and proton
as the particles are assumed to be on-shell.
Such un-physical events are rejected by our code, because they are 
artefacts of the mSPA which corrects for the energy losses due to
photon emission and assumes on-shell vertices.
At some kinematic settings the un-physical events described
above account for a significant fraction of all events, 
changing the total yield.\\

The PA does not have this particular problem since it can fulfil both 
energy and momentum conservation at the same time when assuming massless 
on-shell electrons.
Energy gains through emission of radiation are not possible.
The recoiling protons cannot be assumed to be massless, of course.
But as they only account for a small fraction of high energy
brems\-strah\-lung events, they do not change the total yield
much, neither in the case of the PA nor for the FAMC simulation.

\section{The applicability of the SPA}
\label{sec7}

As we have shown in sect.~\ref{sec2} the SPA
simplifies the multi-photon brems\-strah\-lung treatment considerably.
In order to evaluate its applicability for the kinematic
\begin{figure}
\centering
\includegraphics[height=6cm]{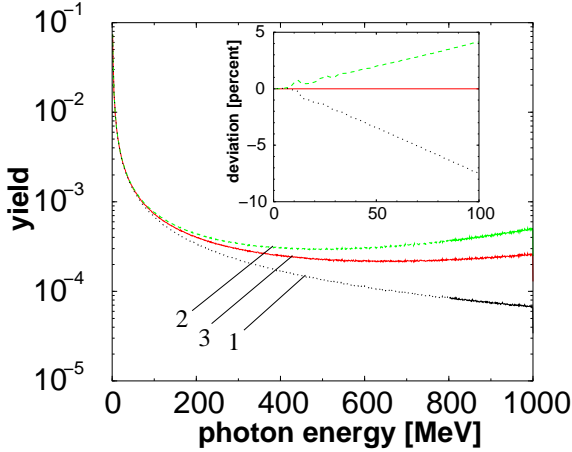}
\caption{\label{fig12} Photon energy distribution for the kinematic
setting shown in table~\ref{tab1}.
The dotted curve (1) (black) represents pSPA where the kinematics used are purely 
elastic.
The dashed curve (2) (green) shows the mSPA calculation which has
been corrected for the kinematic changes due to brems\-strah\-lung.
This mSPA has been used in the present paper and also
in refs.~\cite{makins,simc}.
The solid curve (3) (red) depicts the exact $1\gamma$ calculation for the photon 
energy.
The inset graph shows the difference between the calculations
in percent.}
\end{figure}
\begin{figure}
\centering
\includegraphics[height=6cm]{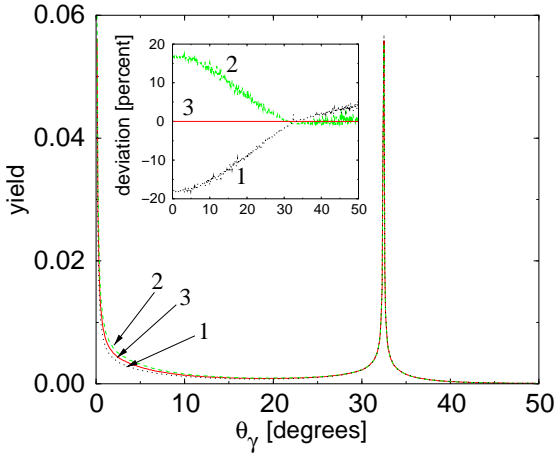}
\caption{\label{fig13} Brems\-strah\-lung angular distribution for
single-photon emission at the kinematic setting shown in table~\ref{tab1}.
The dotted curve (1) (black) represents pSPA, the dashed curve (2) (green) shows the mSPA 
calculation, the solid curve (3) (red) depicts the exact $1\gamma$ calculation 
for the photon
angles.
The inset graph shows the deviation between the different calculations in 
percent.}
\end{figure}
settings considered in this paper we now test the SPA by comparing it to the 
exact $1\gamma$ calculation (omitting proton brems\-strah\-lung).
The integration over the brems\-strah\-lung photons is carried out
with our FAMC generator.
This renders an ana\-ly\-tic evaluation of phase space integrals unnecessary.\\

Let us first describe how we construct a Monte Carlo generator for the
exact $1\gamma$ brems\-strah\-lung calculation from the mSPA Monte Carlo
generator described before.
Consider the SPA brems\-strah\-lung cross section
\begin{eqnarray}
\label{spa1}
\sigma_{\rm SPA}
&\sim&\int\frac{d^3\omega}{2\omega^0}
|{\cal M}_{\rm ep}^{(1)}|^2 A({\omega})\nonumber\\
&=&\int\frac{\omega^0 d\omega^0 d\Omega_\gamma}{2}
|{\cal M}_{\rm ep}^{(1)}|^2 A({\omega}) \, ,
\end{eqnarray}
where we have absorbed the photon energy $\omega^0$ into the
SPA angular distribution,
\begin{eqnarray}
\label{absorbbbb}
A(\omega)\equiv\frac{A(\Omega_\gamma)}{\omega^0} \, .
\end{eqnarray}
Our FAMC code generates brems\-strah\-lung events according to 
the distribution
\begin{eqnarray}
\label{spaA}
\frac{d^3\omega}{2\omega^0} A({\omega})
=\frac{\omega^0 d\omega^0d\Omega_\gamma}{2}A({\omega}) \, .
\end{eqnarray}
Evaluating the phase space integral in Eq.~(\ref{spa1}) using our 
FAMC generator we obtain
\begin{eqnarray}
\label{spa2}
\sigma_{\rm SPA}
&\sim&\frac{1}{N}\sum_{\rm events}
|{\cal M}_{\rm ep}^{(1)}|^2 A({\omega})\frac{\omega^0}{2}
\frac{1}{A({\omega})\frac{\omega^0}{2}}\nonumber\\
&=&\frac{1}{N}\sum_{\rm events}
|{\cal M}_{\rm ep}^{(1)}|^2 \, ,
\end{eqnarray}
where $N$ is the number of events,
and ${\cal M}_{\rm ep}^{(1)}$ is the elastic first-order Born matrix element.
The exact $1\gamma$ calculation (not using the SPA) yields
\begin{eqnarray}
\label{spa3}
\sigma_{1\gamma}
\sim\int\frac{\omega^0 d\omega^0 d\Omega_\gamma}{2}
|{\cal M}_{1\gamma}|^2 \, ,
\end{eqnarray}
where
\begin{eqnarray}
\label{Mex}
{\cal M}_{1\gamma}={\cal M}_{\rm ei} + {\cal M}_{\rm ef} 
\end{eqnarray}
is the exact QED single-photon electron brems\-strah\-lung amplitude.
The cross section (\ref{spa3}) becomes
\begin{eqnarray}
\label{spa4}
\sigma_{1\gamma}
&\approx&\frac{1}{N}\sum_{\rm events}
\frac{|{\cal M}_{1\gamma}|^2}{2}
\frac{1}{\frac{A({\omega})}{2}}\nonumber\\
&=&\frac{1}{N}\sum_{\rm events}
\frac{|{\cal M}_{1\gamma}|^2}{A({\omega})}
\end{eqnarray}
in the Monte Carlo formalism.
In order to measure the applicability of the SPA we assign a weight 
$w_{\rm ex}$ to each event, defined by the ratio of the squared matrix elements
from eqs.~(\ref{spa2}) and (\ref{spa4}), re-weighting our FAMC generator (SPA),
\begin{eqnarray}
\label{spa5}
w_{\rm ex}\equiv
\frac{|{\cal M}_{1\gamma}|^2}{|{\cal M}_{\rm ep}^{(1)}|^2 A(\omega)}\, .
\end{eqnarray}
We compare that weight with the mSPA weight
\begin{eqnarray}
\label{spa6}
w_{\rm mod}\equiv
\frac{|{\cal M}_{\rm ep}^{\rm (1),\,mod}|^2 A^{\rm mod}({\omega})}
{|{\cal M}_{\rm ep}^{(1)}|^2 A(\omega)} \, ,
\end{eqnarray}
where 'mod' indicates that these matrix elements in the numerator
have been calculated for modified kinematics, {\em i.e.}~in mSPA.
Finally we define the trivial weight
\begin{eqnarray}
\label{spa7}
w_{\rm triv}\equiv 
\frac{|{\cal M}_{\rm ep}^{(1)}|^2 A({\omega})}
{|{\cal M}_{\rm ep}^{(1)}|^2 A(\omega)}
= 1 \, ,
\end{eqnarray}
which represents the pSPA, {\em i.e.}~the pure SPA calculation not taking into
account any kinematic changes imposed by emission of brems\-strah\-lung.
This last option is the roughest approximation.
A version of the mSPA represented by weight (\ref{spa6}) is also used in 
ref.~\cite{makins}, combined with the PA.\\

Plotting the photon energy distribution (see fig.~\ref{fig12}) using the
three weights (\ref{spa5}), (\ref{spa6}), and (\ref{spa7}) for each 
Monte Carlo event, we see that the deviation between the
mSPA calculation and the $1\gamma$ calculation for the photon energy is 4.1\% 
for $\omega^0=100\,{\rm MeV}$.
The deviation becomes much lar\-ger for higher energies, 
going up to 90\% for photon energies of $\omega^0=1000\,{\rm MeV}$,
the mSPA calculation \cite{makins,simc} overestimating the radiative tail;
and one can find even larger deviations for different kinematic settings.
However, brems\-strah\-lung events with photon energies of se\-ve\-ral 
hundred MeV are unimportant
for the data analyses since the particle detectors do not see them.
Their momentum acceptances usually are limited to $\pm10\%$ of the
central (elastic) momenta, depending on what one is looking for.\\

The dotted curve (1) in fig.~\ref{fig12} shows the pSPA which was obtained
using weight (\ref{spa7}).
Most data analysis codes make use of some version of the mSPA. 
The mSPA is closer to the exact calculation than pSPA, 
so it constitutes an improvement over pSPA.
This finding is in agreement with ref.~\cite{makins}.\\

For our purposes the influence of the SPA on the photon
{\em angular} distribution is more important than the deviation
in the {\em missing energy} calculation. 
Fig.~\ref{fig13} shows the photon angular distribution. 
The largest deviations occur in the vicinity of the peak due to radiation from
the incident electron, at small values of the angle $\theta_\gamma$.
While the $1\gamma$ calculation of the brems\-strah\-lung
cross section is symmetric in $e$ and $e'$, the mSPA data analysis
procedures are not, resulting in asymmetric deviations from the $1\gamma$ 
result.
This can be understood from the energy loss of the incident electron,
leading to smaller brems\-strah\-lung energies coming from the scattered
electron.
From fig.~\ref{fig1} we know that one critical domain of large discrepancies
between data and standard simulations using the PA is the region in the middle 
between the two radiation peaks where the PA angular distribution 
falls below the measured distribution by a factor 2.
Fig.~\ref{fig13} shows that the FAMC calculation (using mSPA)
overestimates the photon angular distribution in this region by 13\%.
Fig.~\ref{fig6} however suggests, that the FAMC (using mSPA)
reproduces the data well, especially in the region in the middle between 
the two radiation peaks.
Comparing fig.~\ref{fig6} (which includes internal and external 
brems\-strah\-lung, multi-photon emission, finite detector resolution and 
acceptances, multiple scattering and other energy losses) and fig.~\ref{fig13} 
(internal single-photon brems\-strah\-lung only) in the critical region 
between the two electron radiation peaks, we can conclude that the SPA 
impact on the angular distribution is smeared out by other sources of 
inaccuracies like finite detector resolution, finite momentum acceptances, 
and multiple scattering.\\

In fig.~\ref{fig6} we saw that the height of the $e'$ peak is slightly 
underestimated by our FAMC simulation, as well as the large-angle tail
beyond the $e'$ peak.
The results shown in fig.~\ref{fig13} indicate small differences between
the mSPA and the exact $1\gamma$ calculation in the vicinity of the $e'$ peak 
only.
And also the difference in fig.~\ref{fig13} at the
large-angle tail beyond the $e'$ peak is small.
At this stage it is not entirely clear whether removal of the SPA
would affect the photon angular distribution in these regions.

\section{Conclusion and Outlook}
\label{sec8}
Using a FAMC brems\-strah\-lung calculation at almost no extra 
com\-pu\-ta\-tio\-nal ex\-pense im\-proves the treat\-ment
of in\-ter\-nal brems\-strah\-lung in $(e,e'p)$ ex\-pe\-ri\-ments.
One shortcoming of the PA, the un\-der\-esti\-ma\-tion of 
brems\-strah\-lung between the radiation peaks, is solved by our approach.
We have also shown how the PA can be removed.\\

Figs.~\ref{fig7} and \ref{newfig} and table~\ref{tab2} indicated that the FAMC
simulation exposes problems due to the SPA which are hidden when using the PA.
And while for the photon angular distribution the PA may be the dominant 
source of error, we have shown in the previous section that the SPA seems 
to have a sizeable influence on the missing energy distribution.\\

These two problems with the SPA indicate that it would be desirable to also
remove the SPA from data analysis codes.
But there are se\-ve\-ral problems which have to be tackled in order to 
achieve such an improved calculation.
An exact multi-photon calculation would be impracticable since one would 
have to insert the QED cross sections into the analysis codes for 
multi-photon brems\-strah\-lung up to arbitrarily high orders.
One could instead try to combine exact single-photon brems\-strah\-lung
for the hardest brems\-strah\-lung photon with SPA 
multi-photon emission for the softer photons in order to improve the
present brems\-strah\-lung treatment.
Yet, inclusion of proton brems\-strah\-lung, as in the present paper,
seems not to be feasible for such calculations.

\begin{acknowledgement}
The authors are grateful to John Arrington for his support
with the {\sc simc} code. 
And they wish to thank Paul Ulmer and Mark Jones for going through the FAMC
code.
FW also wishes to ac\-know\-ledge the support by the 
{\sc Schwei\-ze\-ri\-sche Na\-tio\-nal\-fonds}.
\end{acknowledgement}

\end{document}